\documentclass[preprintnumbers,nofootinbib]{revtex4}%
\usepackage{eurosym}
\usepackage{graphicx}
\usepackage{amsmath,amssymb}
\usepackage{color}
\usepackage{hyperref}
\usepackage[outdir=./]{epstopdf}
\usepackage{amsmath}
\usepackage{amsfonts}
\usepackage{amssymb}%
\providecommand{\U}[1]{\protect\rule{.1in}{.1in}}
\def\bc{\begin{center}}
\def\ec{\end{center}}

\def\beq{\begin{equation}}
\def\eeq{\end{equation}}

\begin{document}
\title[ ]{Emission cross sections for energetic O$^{+}$($^{4}S,^{2}D,^{2}P$) - N$_{2}$ collisions}
\author{M. R. Gochitashvili$^{1}$}
\author{R. Ya. Kezerashvili$^{2,3}$}
\email{rkezerashvili@citytech.cuny.edu}
\author{D. F. Kuparashvili$^{1}$}
\author{M. Schulz$^{4}$}
\author{N. O. Mosulishvili$^{1}$}
\author{O. G. Taboridze$^{1}$}
\author{R. A. Lomsadze$^{1}$}
\affiliation{\mbox{$^{1}$Department of Exact and Natural Sciences, Tbilisi State University, 0179 Tbilisi, Georgia}
}
\affiliation{\mbox{$^{2}$New York
City College of Technology, The City University of New York, Brooklyn, NY 11201, USA}
}

\affiliation{\mbox{$^{3}$The Graduate School and University Center, \\
The City University of New York, New York, NY 10016, USA }}
\affiliation{\mbox{$^{4}$Missouri University of Science and Technology, Rolla, MO 65409, USA}}

\begin{abstract}
Measurements of emission cross sections for the O$^{+}-$N$_{2}$ collision system
with the incident beam of $1-10$ keV O$^{+}$ in the ground O$^{+}(^{4}S)$ and
metastable O$^{+}(^{2}D)$ and O$^{+}(^{2}P)$ states are reported. The emission cross section
induced by incident ions in the metastable state O$^{+}(^{2}P)$ is much larger
than that for the ground O$^{+}(^{4}S)$ state. The emission cross section of
N$_{2}^{+}$ ion for (0,0), (0,1) and (1,2) bands system is measured and the
ratio of intensities for these bands is established as $10:3:1.$ It is shown
that the cross sections for the N$^{+^{\ast}}$ions emissions in the
dissociative charge exchange processes increase with the increase of the
incident ion energy. The energy dependence of the emission cross section of
the band (0,0) $\lambda=391.4$ nm of the first-negative band system of the
N$_{2}^{+}$ and degree of linear polarization of emission in O$^{+}-$N$_{2}$
collision are measured for the first time. An influence of an admixture of the
ion metastable state on a degree of linear polarization is revealed.
The mechanism of the processes realized during collisions of ground and metastable oxygen ions on
molecular nitrogen have been established. It is
demonstrated that for O$^{+}-$N$_{2}$ collision system the degree of linear
polarization by metastable O$^{+}$($^{2}P$) ions is less compared to those
that are in the ground O$^{+}$($^{4}S$) state and the sign of emission of degree of linear polarization of excited molecular ions does not change.

\end{abstract}
\date{\today}
%\date[Date text]{date}
%\date{\today}
%\date[Date text]{date}
%\date{\today}
%\date[Date text]{date}
%\date{\today}
%\date[Date text]{date}
%\received[Received text]{date}

%\accepted[Accepted text]{date}

%\published[Published text]{date}

\maketitle

\section{Introduction}

The oxygen and nitrogen molecules are two of the most common elements in the
atmosphere and the ionosphere of the Earth. There has been considerable
research in the past 20 years on both the production and loss of vibrationally
excited O$_{2}$ and N$_{2}$ in the thermosphere and their role in ionospheric
processes. Excitation and deexcitation of O$_{2}$ and N$_{2}$ molecules can
occur through direct electron and ion impact, cascade from excited states, and
chemical reactions \cite{1}. Different constituents of the atmosphere have
their characteristic airglow emissions. Atomic oxygen is a very important
constituent of the upper atmosphere. It plays a crucial role in the
atmospheric chemistry at the mesosphere and thermosphere altitudes. Further
more excitation, dissociation and charge transfer between singly charged
O$^{+}$ ions and various atmospheric molecules are relevant to the
low--temperature edge plasma region of current thermonuclear fusion devices
\cite{2}.

The importance of excitation and dissociation processes of O$^{+}-$N$_{2}$ has
been established from the fundamental atomic physics point of view. One
important motivation to study the collision of O$^{+}-$N$_{2}$ pair is the
fact that the cross section is strongly dependent on the initial electronic
state of the O$^{+}$ ions \cite{3,4,5} and, in particular, two low-lying
excited metastable states $^{2}D$ and $^{2}P$ \ of O$^{+}$ have approximate
lifetimes of 3.6 h and 5 s, respectively \cite{6}. As a result of these long
lifetime metastable O$^{+}$($^{2}D$) and O$^{+}$($^{2}P$) states, O$^{+}$ ions
are available for ion -- molecule reactions. The importance of the O$^{+}%
$($^{2}D$, $^{2}P$) metastable states for planetary science have been stressed
for a long time, for instance for Earth \cite{61, 62, 63}, Venus \cite{64},
Europa \cite{65} and Titan \cite{66} atmospheres. In particular, charge
transfer reactions of state-selected atomic oxygen ions O$^{+}$($^{4}S$,
$^{2}D$, $^{2}P$) with nitrogen and oxygen molecules, which are the major
species in the atmosphere, play an important role in atmospheric chemistry.
Various experimental investigations were performed for O$^{+}$ collisions with
H$_{2}$, N$_{2}$, CO, CO$_{2}$, CS$_{2}$ molecules and these studies are
mostly related to charge transfer processes
\cite{7,8,9,10,11,12,13,14,15,16,17,18,19,20,21,22,23,24}. The charge transfer
processes induced by O$^{+}$ collision with N$_{2}$ and O$_{2}$ are important
in the field of aeronomy \cite{25} since the O$^{+}$ is the dominant ion in
the F region \cite{251} of the atmosphere and both metastable species of
O$^{+}$($^{2}D$) and O$^{+}$($^{2}P$) have been detected here too. While the
reactions of O$^{+}$ ions with N$_{2}$ have received a considerable attention
and several experimental studies (see e.g. \cite{26,27}) have been conducted,
there is still some uncertainty related to the magnitude of the total cross
sections of charge transfer and transfer with excitation processes. Since
inelastic processes at low energy strongly depend on the internal electronic
structure of the colliding particles, the presence of metastable excited ions
in the primary ion beam may sometimes significantly influence the observed
cross sections. For example, the study of charge transfers cross sections for
N$^{+}$ and O$^{+}$ ions in collisions with H$_{2}$ molecules and He atoms,
have found that metastable state ions indeed enhance the cross sections,
sometimes by an order of magnitude \cite{28}.

The influence of the excited O$^{+}$($^{2}D$) state is monitored by a charge
transfer reaction in O$^{+}$ + Ar collisions in Ref. \cite{22}. The large
cross section for reactions involving O$^{+}$($^{2}D$) incident ions and small
cross section for reactions of O$^{+}$($^{4}S$) ground state ions is explained
by the energy difference between the O$^{+}$($^{4}S$) recombination energy and
Ar ionization potential for forming Ar$^{+}$($^{2}P_{1/2}$). Collisions of
O$^{+}$($^{2}D$) metastable ions with O$_{2}$ leading to formation of the
O$_{2}^{+}$ ion either in the $A^{2}\Pi_{u}$ or $a^{4}\Pi_{g}$ state involves
only a small energy defect and, as expected, the cross sections for such
processes are found to be quite large. As to the data related to the collision
with N$_{2}$ molecules, the O$^{+}$($^{2}D$) metastable state is also in
resonance with the N$_{2}$($A^{2}\Pi_{u}$) state and the cross section is
expected to be large, while there is 2 eV gap between the O$^{+}$($^{4}S$) and
the nearest N$_{2}^{+}$( $X^{2}\Sigma_{g}^{+}$) state, resulting in a smaller
cross section. Similar considerations also apply to collisions with O$^{+}%
$($^{2}P$). The charge transfer performed with O$^{+}$($^{2}P$) to N$_{2}^{+}%
$($B$, $v=0$) level is near-resonant (energy defect is 0.1 eV) that causes
about a two order of magnitude difference of cross sections between the
metastable O$^{+}$($^{2}P$) state and the ground O$^{+}$($^{4}S$) state
\cite{7}. In \cite{5} it is stated that the abundance of metastable ions is
strongly dependent on the operating conditions. In particular, the abundance
depends on the gas pressure in the ion source and on the energy of impacting
electrons producing ions. There are many options that allow the control of
metastable states and perform measurements with known species
\cite{8,9,10,11,12,13,14,15,16,19}. Translational spectroscopy experiments
have indicated that the long lifetime of $^{2}D$ and $^{2}P$ metastable states
comprise a significant fraction of ions in an O$^{+}$ beam \cite{30}.

The ratio of the metastable electronic states to the ground state can be
varied by a careful control of the energy of the ionizing electrons in the
primary ion source. It is well known (see \textit{e.g. }\cite{32} and
\cite{32A}) that the use of high pressure ion sources can lead to a
significant loss of metastable ions due to collisional quenching mechanisms.
In \cite{32A} the loss of O$^{+}$($^{2}D$) and O$^{+}$($^{2}P$) state ions was
observed in the reaction O$^{+}$($^{2}D$,$^{2}P$) + CO. The relatively low
O$^{+}$($^{2}D$, $^{2}P$) metastable ion abundance was observed in \cite{33},
when a CO pressure of the order of $10^{-1}$ Torr was applied in an ion
source. The beam attenuation method, to determine the fractional abundance of
different states in an O$^{+}$ ion beam formed by the ionization of O$_{2}$,
is considered in an early study \cite{34}. In \cite{35} an appropriate range
of an ion source pressures (0.02$-$0.1 mTorr) was found for the O$^{+}-$
N$_{2}$ collision system and it was reported that the excited metastable
species exhausted by about 3.5 times. As stated by the authors of \cite{35} at
a pressure higher than 2 mTorr the beam can contain either ground state
O$^{+}$ ions or a mix of the ground state and excited-state ions. At a low
pressure of 0.1 mTorr the ion beam is composed of a mix of ground and excited
state ions. The quenching rate of O$^{+}$($^{2}P$) by O and N$_{2}$ have been
evaluated in \cite{36}. In \cite{42} the O$^{+}$ metastable fraction is
estimated to be approximately 0.19 from an ion beam attenuation technique.
Production of the O$^{+}$ ions in \cite{41} expected to be a mix of ground
state and metastable species but again at some unknown ratio. This brief
survey shows that much of the difficulties in the laboratory investigation of
inelastic processes is due to the fact that incident ions are typically
present in both the ground $^{4}S$ state and in the long-lived excited
metastable $^{2}D$ and $^{2}P$ states. It should be noted that metastable
species are attenuated differently and that the composition of the total
metastable fraction is different. Furthermore, it seems that the data related
to the abundance for different targets are contradictory. This difference is
also observed for the same target case. The source of this large discrepancy
between an abundance of metastable state can be attributed to a different
operating condition of ion sources. We can conclude that the magnitudes of
cross sections for inelastic processes are largely dependent on the
composition of the beam. Thus, for reactions involving chemically stable
molecular targets, the experimental difficulties are primarily associated with
the determination of the composition of the primary ion beam. Most of the
excited-state measurements reported generally pertain to an undetermined
mixture of O$^{+}$($^{2}D$) and O$^{+}$($^{2}P$) ions and in the 0.2--10 keV
range, the uncertainty is on the order of $\pm$35 \% and it is now known that
the cross sections for these two excited-state species differ \cite{421}.
Finally let us mention that the charge transfer processes of O$^{+}$ with
N$_{2}$ for collision energies from 10 eV to 100 keV are reviewed by Lindsay
and Stebbings in Ref. \cite{421}, which also gives a brief descriptions of the
experimental measurement techniques. To control the metastable species in the
ion beam both the precise exploration of pressure condition in the ion source
and the evaluation of electrons energy used in the ion source for ionization
are crucial.

In this paper we study the emission processes in collisions of the ground
O$^{+}(^{4}S)$ and long lifetime metastable O$^{+}(^{2}D)$ and O$^{+}(^{2}P)$
state ions with N$_{2}$ molecule in the ion energy range 1--10 keV. The paper
is organized in the following way. In Sec. II the method and procedure of
measurements using an optical spectroscopy are presented. Here we introduce
our approach and procedure for cross section measurements and its
determination. The energy dependence of the emission cross sections and degree
of linear polarization for O$^{+}-$N$_{2}$ processes induced by the incident
beam of O$^{+}$ in the ground and metastable states and discussion of results
of measurements are presented in Sec. III. Finally, in Sec. IV we summarize
our investigations and present conclusions.

\section{METHOD AND PROCEDURE OF MEASUREMENTS}

\bigskip The experimental study of emission processes in reactions of oxygen
ions in the ground and metastable state with nitrogen molecules is performed
within the framework of optical spectroscopy.
%First, to investigate the formation of charged ions we use electron impact to produce charged ions by using electron spectroscopy. To calibrate and check the mass transmission we measure the electron induced ionization of CH$_{4}$ molecules and Ar atoms. Second, we study the ion production in $e+$O$_{2}$ collision as a function of the incident electron
%energy for different pressure conditions of the molecular oxygen in the ion
%source. These investigations allow us to find conditions for reliable beams of the ground O$^{+}$($^{4}S$) and metastable O$^{+}$($^{2}D$) and O$^{+}$% ($^{2}P$) ions to study the excitation processes induced by these ions in collisions with N$_{2}$ molecules. The latter measurement we perform using optical spectroscopy.
This approach allow us to study the emission processes induced by reliable
beams of the ground O$^{+}$($^{4}S$) and metastable O$^{+}$($^{2}D$) and
O$^{+}$($^{2}P$) ions in collisions with N$_{2}$ molecules.

In our study we used a radio frequency (RF) discharge ion source for the
investigation of the energy dependence of the emission cross section in
collisions of oxygen ions with nitrogen molecules. Optical spectroscopy, used
to study the energy dependence of the emission cross section, is characterized
by a high energy resolution, but low luminosity, since radiation is observed
only at a certain angle and for a certain spectral line. Therefore, it is
necessary to study the process for a relatively large current of O$^{+}$ ions.
The radio frequency source allows to obtain much higher ion currents ($\sim$
%TCIMACRO{\U{3bc}}%
%BeginExpansion
$\mu$%
%EndExpansion
A). Our experiments showed that, depending on the condition (pressure, power,
gas mixture) of ion beam formation, the fraction of O$^{+}$ ions in different
electronic states (ions in $^{4}S$ and metastable $^{2}D$, $^{2}P$ states)
significantly changes, which significantly affects the dependence of the
emission cross section on the ion impact energy. Establishing the formation
mechanism of the ion beam in the RF source, one faces challenges. When the
pressure inside the source changes, due to various inelastic processes (charge
exchange process, excitation, quenching, recombination), the ion filtration
process takes place. In this case, the degree of filtration and neutralization
of ions depends on the internal electronic state of the ion. An increase in
pressure decreases the free path of electrons, which participate in the
formation of ions in various electronic states via the process of dissociative
ionization. Consequently, the probability of formation of ions in different
states changes. In order to determine the mechanism of the formation of
O$^{+}$ ions, we investigated the dependence of the dissociative ionization
cross section on the electron energy inside the ion source for various
pressures of the working gas.

\subsection{Optical Spectroscopy}

Cross sections for the emission processes were measured by optical
spectroscopy. We have used an experimental setup and method of measurements
similar to those described in detail and used previously in \cite{43}. To
obtain the oxygen beam we used a 20 MHz RF ion source. The oxygen ions
extracted from the RF ion source, are accelerated, collimated and focused, and
mass-selected by a 60$^{0}$ magnetic sector field. Then the ion beam is
directed into the collision chamber. The fluorescence emitted as a result of
the excitation of colliding particles was observed at 90$^{0}$ with respect to
the beam. In the present work, measurements are performed with sufficiently
high spectral resolution, which allows to distinguish the emission channels.
This method also allowed us to estimate the polarization of emission, that
itself is a powerful tool for establishing the mechanism for inelastic
processes. The spectroscopic analysis of the emission is performed with a
monochromator incorporating a diffraction grating with a resolution of 40
nm/mm operated in the $400-800$ nm spectral region, which allows to
distinguish the excitation channels. A polarizer and a mica quarter-wave phase
plate are placed in front of the entrance slit of the monochromator and the
linear polarization of the emission is analyzed. For cancelation of the
polarizing effect of the monochromator, the phase plate is placed after the
polarizer and is rigidly coupled to it. The emission is recorded by a
photomultiplier with a cooled cathode and operated in current mode. The
spectral sensitivity calibration is performed by a tungsten filament standard
lamp, which is chosen due to the lack of reliable experimental data in the
infrared region (the bright Meinel system), that could be used for the
calibration of the system for registration of radiation \cite{432}. The
measurements for low energy collisions required a precise determination of the
energy of the ions as well as their energy dispersion. To avoid errors in the
measurements of the incident ion energy we employ the retarding potential
method and use the electrostatic analyzer with a resolving power of 500. In
our experiments the energy dispersion of the ions does not exceed 20 eV. The
ion current in the collision chamber was of the order 1$-$5 $\mu$A. The
residual gas pressure did not exceed $10^{-6}$ Torr. The working pressure in
the collision chamber was about $5\times10^{-4}$ Torr. Single collision
condition was checked by a linear dependence of the intensity of spectral
lines versus target gas pressure and density of the ion current. The absolute
accuracy of the measurements is 25\%. The accuracy of the measurements is
related to the following factors: i. the accuracy of the pressure measurements
in the collision chamber; ii. the precise and immediate determination of the
primary beam current that produce the radiation, which is collected by optical
system from the region where the collision occurs; iii. the accuracies of the
relative and absolute calibration procedures. The uncertainty of the relative
measurements is about 5\%.

\subsection{Cross section measurements and determination}

\begin{figure}[b]
\centering
\includegraphics[width=8.0cm]{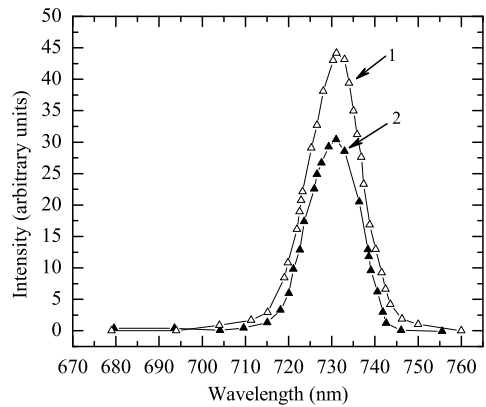}\caption{The typical spectra of
emission of Ar atomic line ($\lambda=731.1$ nm) for O$^{+}-$Ar collision
system at fixed ($E=2.5$ keV) energy of O$^{+}$ colliding ions and for two
different gas pressure in the RF ion source. Curves: 1 - for the pressure
$P=1.5\times10^{-2}$ Torr; 2 - for the pressure $P=1\times10^{-1}$ Torr,
respectively, in the RF ion source. }%
\label{Fig9}%
\end{figure}

One advantage of the RF source used in this experiment is that, depending on
its operating mode, a certain fraction of the ions can be found in metastable
states and easily identified. As it was mentioned in the introduction the
probability of quenching is significantly related to the increase of the
target pressure in the ion source. Hence, it is important to explore the
working condition of a source by changing the pressure in it. Therefore we
considered the following reaction%

\begin{equation}
\text{O}^{+}+\text{Ar}\rightarrow\text{O}^{+}+\text{Ar}^{\ast}(6s).
\label{Argon}%
\end{equation}
A typical emission spectra for the intense spectral line of Ar atom
($\lambda=731.1$ nm) in O$^{+}-$Ar collision for the process (\ref{Argon}) at
different pressures in the ion source are presented in figure \ref{Fig9}. It
was found that by increasing the O$_{2}$ gas pressure in the ion source about
6 times (from $1.5\times10^{-2}$ to $1.0\times10^{-1}$ Torr) the intensity of
this lines decreases by a factor of 1.4. This result indicates that the
presence of metastable ions plays a definite role in excitation of the
Ar(6$s$) state (the excitation energy is $14.85$ eV) and hence, this process
is effectively realized due to the transfer of internal energy. In order to
check the influence of the ion source operating parameters and to evaluate the
role of metastable ions in the excitation processes we have extended this
research to nitrogen molecules and to its dissociative products. For the
latter we chose such energy and pressure values that the observation of a
metastable state could be possible at high energy and for a single collision condition.

\begin{figure}[b]
\centering
\centering
\par
\includegraphics[width=7.9cm]{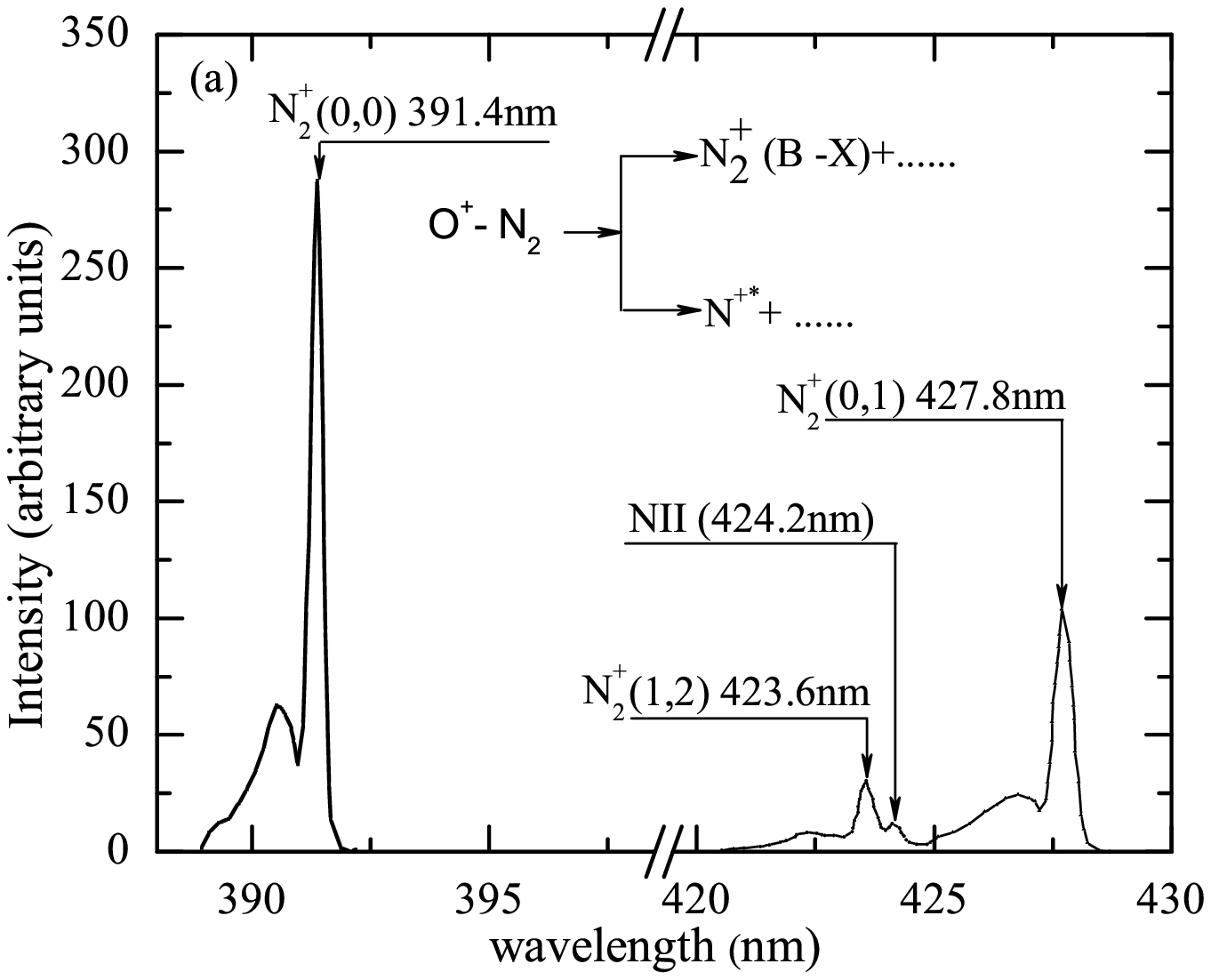}
\includegraphics[width=7.9cm]{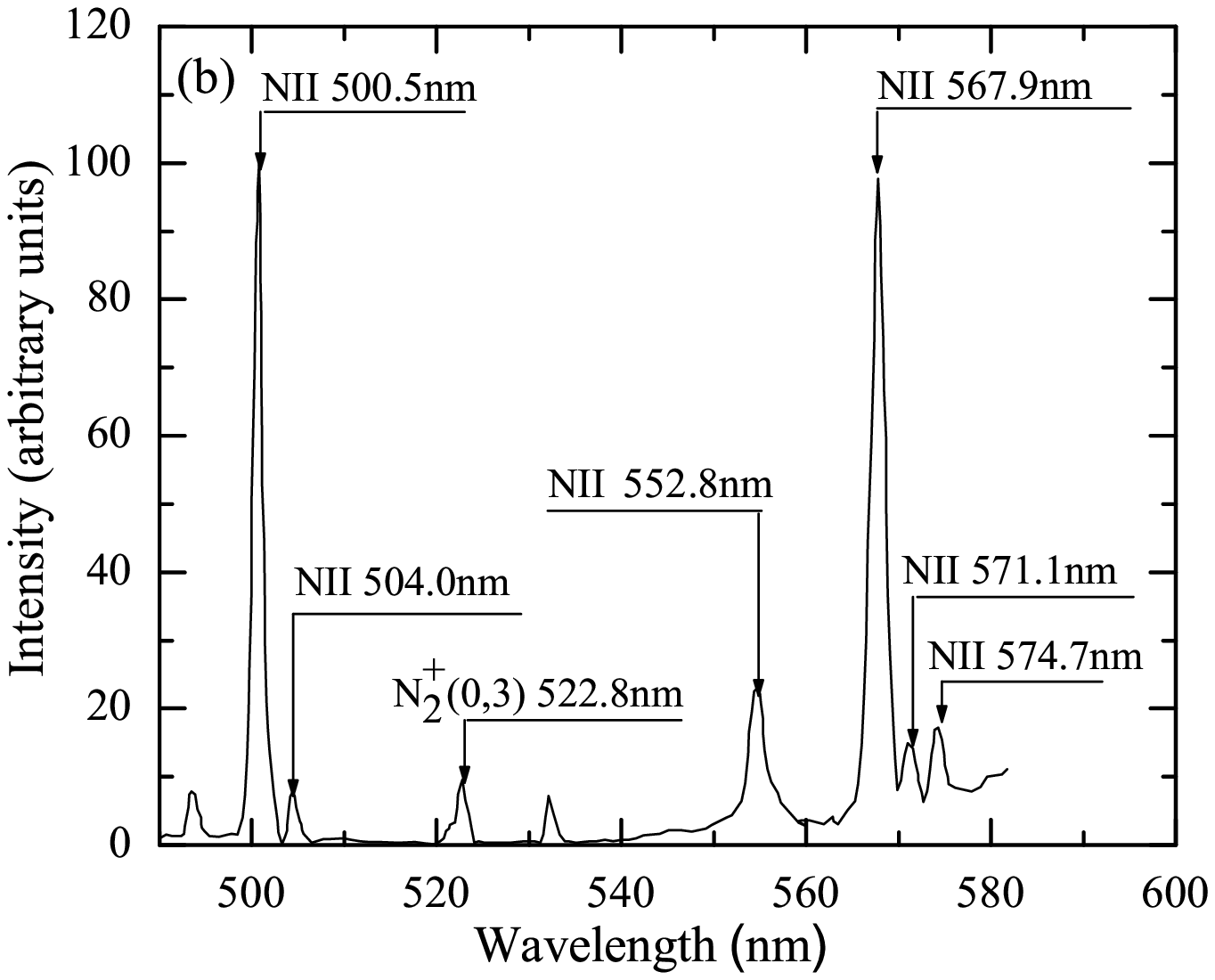}\caption{ Intensity of the emission
for the bands of N$_{2}^{+}$ ion\ for transitions N$_{2}^{+}B^{2}\Sigma
_{u}^{+}(v^{\prime})\rightarrow$N$_{2}^{+}X^{2}{}_{g}^{+}(v^{\prime\prime}%
)$, where $v^{\prime}=0$ or  $v^{\prime}=1$ and $v^{\prime\prime}=0$,1,2,3
and dissociative product of N$^{+}$ ionic lines in O$^{+}-$N$_{2}$ collision
at the energy $E=5$ keV and pressure in the RF ion source $P=7\times10^{-3}$
Torr versus the wavelength. Measured wavelengths are as labeled on the figure.
(a) (0,0), (0,1), and (1,2) bands of the N$_{2}^{+}$ ion and NII 424.2 nm line
of the N$^{+}$ ion. (b) (0,3) band of the N$_{2}^{+}$ ion and NII 500.5 nm,
NII 504.0 nm, NII 552.8 nm, NII 567.9 nm, NII 571.1 nm and NII 574.7 nm lines
of the N$^{+}$. }%
\label{Fig6}%
\end{figure}

We performed a test experiment for the intensity of the emission of the (0,0),
(0,1) and (1,2) bands of molecular N$_{2}^{+}$ ions and of the dissociative
N$^{+}$ fragment in collision of the O$^{+}$ ion with N$_{2}$ molecules. The
intensity of emission depends on the projectile ion energy. A typical spectrum
at the energy $E=5$ keV of the primary beam and pressure $P=7\times10^{-3}$
Torr in the ion source is shown in figure \ref{Fig6}. As it is seen from
figure \ref{Fig6}a the ratio between the intensities of these bands is
$10:3:1$, respectively. As the relative intensity has been established, we
chose an intensive band (0,0) N$_{2}^{+}$ ion ($\lambda=391.4$ nm) and
performed a study with variable pressure. Such procedure allowed us to
separate the ground and metastable state ions in the primary beam and to
perform a systematic measurement. In the same figure \ref{Fig6}a the emission
of the satellite dissociative product for the nitrogen ionic line,
$\lambda=424.2$ nm, is also presented, while in figure \ref{Fig6}b the band
(0,3) of the N$_{2}^{+}$ ion emission, $\lambda=522.8$ nm, is shown.

The main objective of our research is the study of the emission processes in
reactions of the ground O$^{+}(^{4}S)$ and metastable O$^{+}(^{2}D)$ and
O$^{+}(^{2}P)$ state ions with N$_{2}(X^{1}\Sigma_{g}^{+})(v^{\prime}=0)$
molecule in a wide energy range of incident ions. In particular, in collisions
of O$^{+}$ ions with nitrogen molecules the studies are carried out for the
following inelastic channels with formation of the excited state $B^{2}%
\Sigma_{u}^{+}(v^{\prime}=0$ or $v^{\prime}=1)$ of the nitrogen
molecular ion and the dissociation products of N$^{+}$.%

\begin{align}
\text{O}^{+}\text{(}^{4}S\text{)}+\text{N}_{2}(X^{1}\Sigma_{g}^{+}%
)(v^{\prime} &  =0)\rightarrow\text{N}_{2}^{+}(B^{2}\Sigma_{u}%
^{+})(v^{\prime}=0\text{ or }v^{\prime}=1)+...\\
\text{O}^{+}\text{(}^{2}D+^{2}P\text{)}+\text{N}_{2}(X^{1}\Sigma_{g}%
^{+})(v^{\prime} &  =0)\rightarrow\text{N}_{2}^{+}(B^{2}\Sigma_{u}%
^{+})(v^{\prime}=0\text{ or }v^{\prime}=1)+...\\
\text{O}^{+}\text{(}^{4}S\text{)}+\text{N}_{2}(X^{1}\Sigma_{g}^{+}%
)(v^{\prime} &  =0)\rightarrow\text{N}^{+^{\ast}}+\text{N}+\text{O}%
(^{3}P)\\
\text{ \ \ \ } &  \downarrow\nonumber\\
&  \rightarrow\text{N}^{+^{\ast}}(3d\text{ }^{3}F^{0})+\text{N}+\text{O}%
(^{3}P)\\
&  \rightarrow\text{N}^{+^{\ast}}(3p\text{ }^{3}D)+\text{N}+\text{O}(^{3}P)\\
&  \rightarrow\text{N}^{+^{\ast}}(4fF)+\text{N}+\text{O}(^{3}P)\\
\text{O}^{+}\text{(}^{2}D+^{2}P\text{)}+\text{N}_{2}(X^{1}\Sigma_{g}%
^{+})(v^{\prime} &  =0)\rightarrow\text{N}^{+^{\ast}}+\text{N}%
+\text{O}(^{3}P)\\
&  \downarrow\nonumber\\
&  \rightarrow\text{N}^{+^{\ast}}(3d\text{ }^{3}F^{0})+\text{N}+\text{O}%
(^{3}P)\\
&  \rightarrow\text{N}^{+^{\ast}}(3p\text{ }^{3}D)+\text{N}+\text{O}(^{3}P)\\
&  \rightarrow\text{N}^{+^{\ast}}(4fF)+\text{N}+\text{O}(^{3}P)
\end{align}

In the reactions (4) and (8) the dissociative product N$^{+^{\ast}}$ can be in
different excited states, (5)--(7) and (9)--(11), respectively. For the molecule N$_{2}$ in reactions (2)--(4) and (8) the  vibrational quantum number is $v^{\prime}=0$, and for the excited molecular ion N$_{2}^{+}$ in reaction (2) and (3) the vibrational quantum number can be either $v^{\prime}=0$ or $v^{\prime}=1$.

\section{RESULTS OF MEASUREMENTS AND DISCUSSION}

\begin{figure}[b]
\centering
\includegraphics[width=10.0cm]{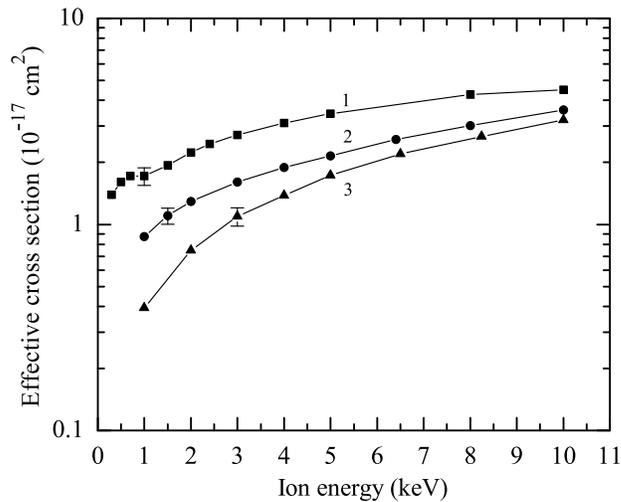}\caption{ Energy dependence of the
effective cross section for the band (0,0) $\lambda=391.4$ nm of the first-negative band system of N$_{2}^{+}$ ion in O$^{+}-$N$_{2}$ collision for different
pressure in the RF ion source. Curves: 1 - for the pressure $P=2.4\times
10^{-2}$; 2 - $P=8\times10^{-2}$; 3 - $P=2\times10^{-1}$ Torr, respectively,
in the RF ion source.}%
\label{Fig7}%
\end{figure}

\bigskip The characteristic emission of N$_{2}$ and N$_{2}^{+}$ are generally
attributed to the first-positive $(B$ $^{3}\Pi_{g}^{+}\rightarrow A$
$^{3}\Sigma_{u}^{+})$ and first-negative $B^{2}\Sigma_{u}^{+}$ $\rightarrow$ $X^{2}%
\Sigma_{g}^{+}$ band systems of N$_{2}$ and N$_{2}^{+}$,
respectively \cite{FLAGAN}. In collisions of O$^{+}$ ions with N$_{2}$, we studied the emission
of the band (0,0) of the first-negative band system of the N$_{2}^{+}$ which
corresponds to the transition $B^{2}\Sigma_{u}^{+}$ $\rightarrow$ $X^{2}%
\Sigma_{g}^{+}$. The study is carried out with O$^{+}$ in the ground state
$^{4}S$ and metastable states ($^{2}D,$ $^{2}P$). The presence of ions in
metastable states is monitored by measuring the energy dependence of the
emission cross section of the (0,0) $\lambda=391.4$ nm bands of the first-negative band system of N$_{2}^{+}$ ion for different pressure conditions of the RF
source. The energy dependence of the effective emission cross section of
N$_{2}^{+}$ ion in collisions of the oxygen ion with nitrogen molecules at
different pressures in the RF source, in particular, at $P_{1}=2.4\times
10^{-2}$ Torr, $P_{2}=$ $8\times10^{-2}$ Torr, and $P_{3}=2\times10^{-1}$
Torr, are presented in figure \ref{Fig7}. The curves 1, 2, and 3 in figure
\ref{Fig7} show our results for the effective cross section obtained for the
corresponding pressures in the RF source. The effective emission cross section
refers to the sum of terms, each of which is the product of the emission cross
section of the given band by ions in the corresponding excited state and the
relative weight of these ions in the primary particle beam. In general, the
value of the effective cross section for O$^{+}-$N$_{2}$ collision can be
written as%

\begin{equation}
\sigma_{eff}(P,E)=f_{0}(P)\sigma_{0}(E)+f_{1}(P)\sigma_{1}(E)+f_{2}%
(P)\sigma_{2}(E), \label{Effective}%
\end{equation}
where $\sigma_{0}(E),$ $\sigma_{1}(E)$ and $\sigma_{2}(E)$ are the
corresponding emission cross sections induced by the ground O$^{+}$($^{4}S$)
and metastable O$^{+}$($^{2}D$) and O$^{+}$($^{2}P$) ions of the primary
O$^{+}$ion beam, which depend on the energy of the beam, while $f_{0}(P)$,
$f_{1}(P)$ and $f_{2}(P)$ are the corresponding relative weight of the ground
O$^{+}$($^{4}S$) and metastable O$^{+}$($^{2}D$) and O$^{+}$($^{2}P$) ions in
the incident O$^{+}$ion beam at the given pressure in the collision chamber.
The relative weights depend on the pressure in the collision chamber and at
the same time $f_{0}(P)+f_{1}(P)+f_{2}(P)=1$.

\begin{figure}[t]
\centering
\includegraphics[width=10.0cm]{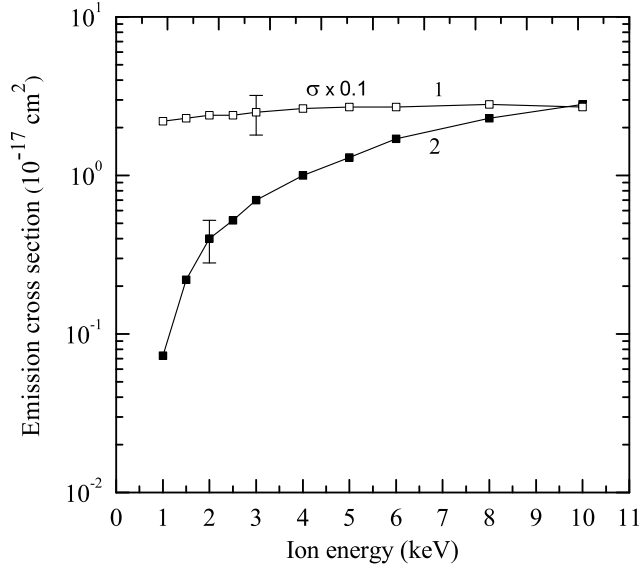}\caption{ Energy dependence of the
emission cross section for the (0,0) $\lambda=391.4$ nm band of the first-negative band system of N$^{+}$${_{2}}$ ion in O$^{+}-$N$_{2}$ collision. Curves: 1
- measurements for O$^{+}$ ion in $^{2}$P metastable state; 2 - measurements
for O$^{+}$ ion in $^{4}$S ground state. }%
\label{Fig8}%
\end{figure}

\begin{figure}[t]
\centering
\includegraphics[width=10.0cm]{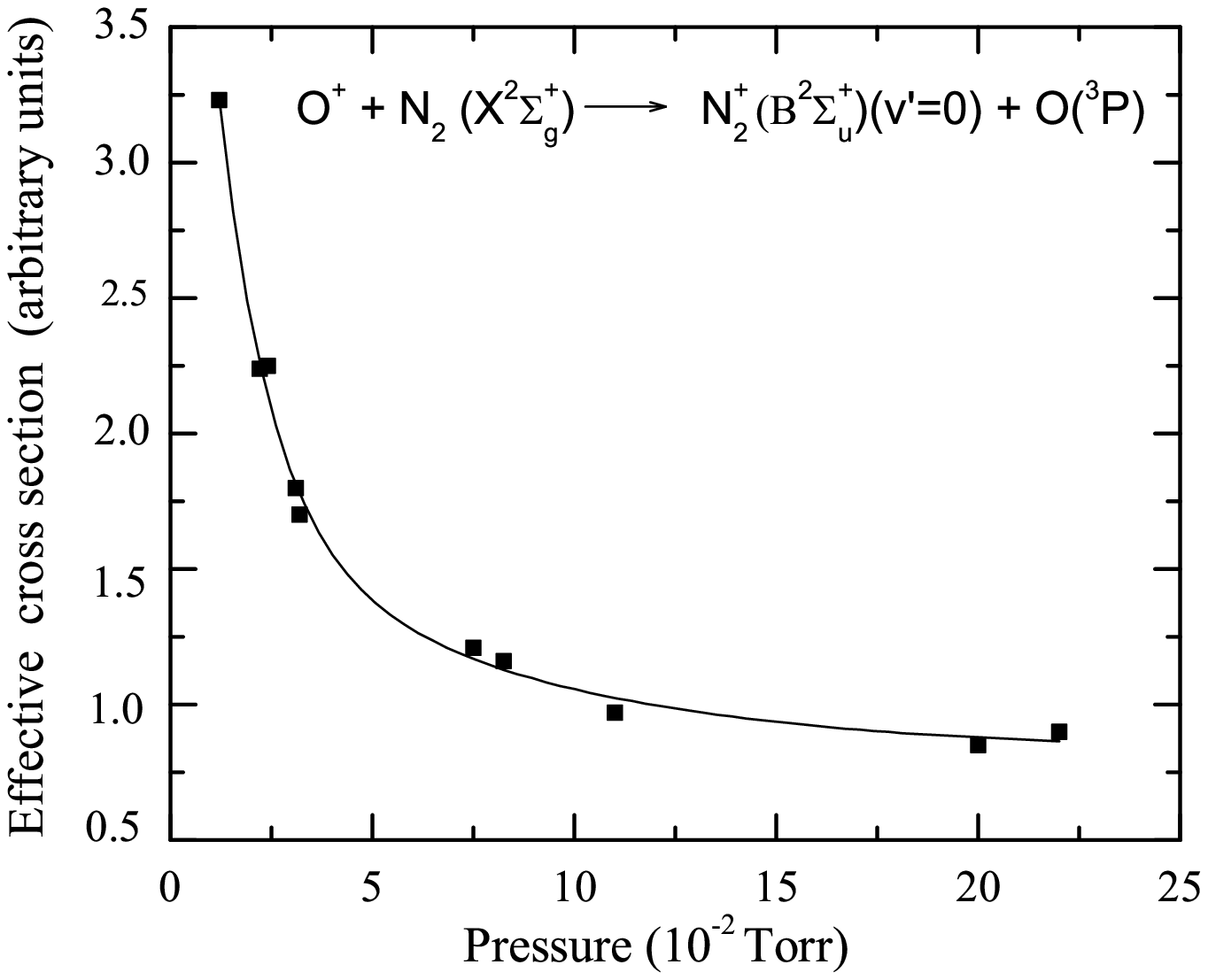}\caption{ Dependence of the effective
cross section for the (0,0) $\lambda=391.4$ nm band of the first-negative band system of N$_{2}^{+}$ ion in O$^{+}-$N$_{2}$ collision on the pressure in the
RF ion source at the incident ions energy $E=2.5$ keV. }%
\label{Fig10}%
\end{figure}

It is known, that the excitation of the indicated band system by oxygen ions
in the $^{4}S$ ground state proceeds with a significantly lower efficiency
than excitation by ions in the metastable $^{2}D$ and $^{2}P$ states. The
ratio of the corresponding cross sections is $1:9:300$ \cite{7}. As can be seen from the graph in figure \ref{Fig7}, there is the difference
in the magnitudes for the effective cross sections at the different pressure, which depends on the incident energy of O$^{+}$ ions. At a fixed energy, the effective cross section decreases with
increasing pressure in the RF source. The absolute emission cross section
%for oxygen ions in the $^{4}S$ ground state and the metastable $^{2}D$ and $^{2}P$ states
for the energy of the incident beam $E=1$ keV obtained in \cite{7} are
$\sigma_{0}=0.73\times10^{-18}$ cm$^{2},$ $\sigma_{1}=6.6\times10^{-18}$
cm$^{2},$ $\sigma_{2}=2.20\times10^{-16}$ cm$^{2}$, which correspond to
$^{4}S$ , $^{2}D$ and $^{2}P$ states of the O$^{+}$ ions, respectively. The
comparison of these results with the effective cross section $\sigma
_{eff}=1.7\times10^{-17}$cm$^{2}$ that we measured at the same energy and
pressure $P_{1}=2.4\times10^{-2}$ Torr in the RF source shows, that our result
is greater than the emission cross sections $\sigma_{0}$ and $\sigma_{1}$ for
the oxygen ions in the $^{4}S$ and $^{2}D$ states and significantly less than
for the O$^{+}$ ions in the metastable $^{2}P$ state obtained in \cite{7}.
This suggests that the value of the effective cross section we obtained is
determined to a significant extent by oxygen ions in the metastable $^{2}P$ state.

At the next step we investigated the presence of the metastable $^{2}D$ state
in the oxygen ions beam, by examining the emission of the Meinel bands (3,0)
$\lambda=687.4$ nm and (4,1) $\lambda=703.7$ nm, which correspond to the
transition $A^{2}\Pi_{u}-X^{2}\Sigma_{g}^{+}$ of the N$_{2}^{+}$ ion in
O$^{+}-$N$_{2}$ collisions. The experiments showed that, despite the
quasiresonant character of these processes (energy defect $\Delta E=0.06$ eV)
the Meinel bands are excited with a low probability, which is probably due to
the absence of the oxygen ions in metastable $^{2}D$ state in the primary
particle beam. The small fraction of these ions in the primary particle beam
in comparison with O$^{+}$($^{2}P$) ions is due to a peculiarity of the
mechanism of formation of these ions during the dissociative ionization in
collisions of electrons with oxygen molecules in the RF ion source. An
analysis of the energy terms of the oxygen molecule and the oxygen molecular
ion shows that the most efficient mechanism of formation of dissociation
products in various states is a decay of highly excited states of the
O$_{2}^{+}$ ion in the region of internuclear distances corresponding to the
Franck--Condon transitions. The same result follows from Ref. \cite{48}. In
this paper is shown that during electron impact dissociative ionization of
O$_{2}$ molecules mainly ions in the ground O$^{+}$($^{4}S$) state and
metastable O$^{+}$($^{2}P$) state are created. It means that the probability
for formation of O$^{+}$($^{2}P$) is higher compared to O$^{+}$($^{2}D$).

\bigskip\ As our measurement shows, as well as its follows from \cite{7}, the
formation of the metastable O$^{+}$($^{2}D$) ions has low probability. Hence
we can infer that in Eq. (\ref{Effective}) term $f_{1}(P)\sigma_{1}%
(E)\approx0$ and, therefore,\textit{ }$\sigma_{eff}=f_{0}(P)\sigma
_{0}(E)+f_{2}(P)\sigma_{2}(E),$ where now $f_{0}(P)+f_{2}(P)=1$\textit{.}
Using these equations one can estimate the value of $f_{0}(P)$ and $f_{2}%
(P)$\ for the given $\sigma_{0}(E)$ and $\sigma_{2}(E)$ and the different
pressure conditions. Let us estimate the value of $f_{0}(P)$ and $f_{2}%
(P)$\ at a fixed energy, $E=1$ keV, and for the different pressure conditions.
Using the values of the effective cross sections $\sigma_{eff}$ measured at
$E=1$ keV and different pressures, and the values of the $\sigma
_{0}=0.73\times10^{-18}$ cm$^{2}$ and $\sigma_{2}=2.2\times10^{-16}$ cm$^{2}$
at $E=1$ keV from \cite{7}, by solving the above mentioned equations one
obtains: $f_{0}=0.926$ and $f_{2}=0.074$ for $P_{1}=2.4\times10^{-2}$ Torr$,$
$f_{0}=0.966$ and $f_{2}=0.034,$ for $P_{2}=8.0\times10^{-2}$ Torr, and
$f_{0}=0.986$ and $f_{2}=0.014$ for $P_{3}=20.0\times10^{-2}$ Torr. Thus, the
portion of metastable ions $f_{2}$ decreases, while the portion $f_{0}$ of
ground state ions increases with increasing pressure in the ion source. It
should be noted that a main reason for the attenuation of the ion beam is
caused by charge exchange processes in the ion source.
%Therefore, the proportion of relative weights of the metastable O$^{+}$($^{4}S$) and O$^{+}$($^{2}P$) states is weakly depends on the pressure and is almost the same.
If we assume, that the relative weights $f_{0}$ and $f_{2}$ are functions of
pressure and do not depend on the energy, then it is possible to determine the
emission cross section for different energies of incident ions being in the
ground O$^{+}(^{4}S)$ and metastable O$^{+}(^{2}P)$ states. This assumption is
true because the energy of emerging ions from the high frequency ion source is
the same (extraction potential of the ions and potential of a plasma in the
discharge area is a constant) and energy of ions is defined in a focusing area
e.g. outside of ion source. It should be mentioned, that the relative weight
of metastable O$^{+}(^{2}P)$ ions, in effective cross section, is significant
at relatively small energies and increases for the ground state O$^{+}(^{4}S)$
ions with the increase of energy.
%This fact is confirmed by our measurements presented in figure \ref{Fig7}.

Since the relation between the relative weight for different pressure
conditions is established, we can use this ratio of relative weight for the
effective cross section at a different energy. For a particular incident
energy we can write a simple system of equation by solving of which one can
restore the energy dependence of the absolute emission cross section as for
the ground O$^{+}(^{4}S)$ state and metastable O$^{+}(^{2}P)$ state as well.
The results of this reconstruction are shown in figure \ref{Fig8}.

The assumption regarding the fact that the ratio of a portion for metastable
and ground state ions does not depend on the collision energy can be confirmed
from our results shown on figure \ref{Fig7}. The analyses of the results shows
that by increasing the pressure the portion of metastable ion beam decreases,
hence the effective emission cross section decreases as well. At the same
time, we know that $\sigma_{2}\gg\sigma_{1}\gg\sigma_{0}$. Here $\sigma_{0},$
$\sigma_{1},$ and $\sigma_{2}$. stands for $^{4}S,$ $^{2}D$ and $^{2}P$ cross
sections, respectively. So, if we take the difference between cross section
for two different pressure condition 1) $\sigma_{eff}(P_{1},E)-\sigma
_{eff}(P_{2},E)$ and $\sigma_{eff}(P_{1},E)-\sigma_{eff}(P_{3},E)$ and for a
different collision energy and normalized them to each other we will get the
energy dependence of the emission cross section for metastable states in
arbitrary units. The results can be calibrated on the absolute cross section
from \cite{7} and we will get a curve which within the error bar (10\%)
coincides with the results shown in Fig. 4. For example, our estimation shows,
that the difference between the effective emission cross section for $P_{1}$
and $P_{3}$ pressure $\sigma_{eff}(P_{1},E)-\sigma_{eff}(P_{3},E)$ as shown in
figure \ref{Fig8}, in the energy interval $1-10$ keV makes 6 $\%$ of total
cross section of metastable particles. This value is in good agreement with
our previous estimation (7.4\%) for the total metastable portion which we did
for the energy $E=1$ keV. From this result also follows, that the remaining
1.4\% \ ($7.4\%-6\%$) in the case of pressure condition $P_{3}=2.0\times
10^{-1}$ Torr in the ion source, is still related to the presence of
metastable states in ion current. Taking this result into consideration we can
obtain the energy dependence of the absolute emission cross section for the
ground $^{4}S$ state ion.
%This result is taking in to consideration in figure \ref{Fig8}.
This also justified by the fact that when the pressure in the ion source is
$P_{3}=2.0\times10^{-1}$ Torr the weight of the metastable ion is about 1.4 \%
in total ion beam, in all energy interval.
%Take all this into consideration, once again we can obtain the final results for energy dependence of cross sections, which are presented in figure \ref{Fig8}.
It should be mentioned, that an accuracy of absolute cross section, in low
energy is about 30\% and the accuracy does not exceed 15 \% for high energy.
Using the data from \cite{7}, we can estimate from our results the fraction of
oxygen ions in the $^{4}S$ ground state and in the metastable $^{2}P$ state in
the primary particle beam for different values of the pressure of the working
gas in the RF source. Simple calculations show that the fraction of O$^{+}%
$($^{2}P$) ions in the primary particle beam for the pressures of
$2.4\times10^{-2}$ Torr, $8\times10^{-2}$ Torr and $2\times10^{-1}$ Torr in
the RF source is 7.4\%, 3.4\% and 1.4\%, respectively. As we mentioned our
measurements show that the O$^{+}$($^{2}P$) ions make a substantial
contribution to the effective cross section. Therefore, the decrease in the
effective cross section resulting by the increase of the pressure inside the
RF source is entirely due to the decrease of the fraction of these ions in the
primary beam. The contribution of the metastable ions to the effective cross
section also depends substantially on their energy --- it decreases as their
energy increases. After determining the relative weight of metastable ions in
the primary beam, it is easy to reconstruct the dependence of the cross
section of emission of the (0,0) $\lambda=391.4$ nm band by O$^{+}$ ions in
the $^{4}S$ ground state and in the metastable $^{2}P$ state from the energy
dependence of the effective emission cross section by solving some simple
algebraic equations. The results of this reconstruction are shown in figure
\ref{Fig8}. It can be seen that the cross section for emission of the given
band by O$^{+}$($^{2}P$) and O$^{+}$($^{4}S$) ions behave differently. The
emission cross section for the incident O$^{+}$ ion in $^{2}P$ state slightly
increases in the energy range $1-5$ keV and is almost constant in the energy
range $5-10$ keV. The emission cross section for the ground state incident
O$^{+}$ ion $^{4}S,$ sharply increases in the entire energy range of the
incident beam, and at higher ion energies the contribution of the ions found
in the $^{4}S$ ground state to the excitation becomes decisive. However, the
magnitude of this cross section is still more than one order of magnitude
smaller than the one for O$^{+}(^{2}P)$ ions at the energy 10 keV.

\begin{figure}[t]
\centering
\includegraphics[width=10.5cm]{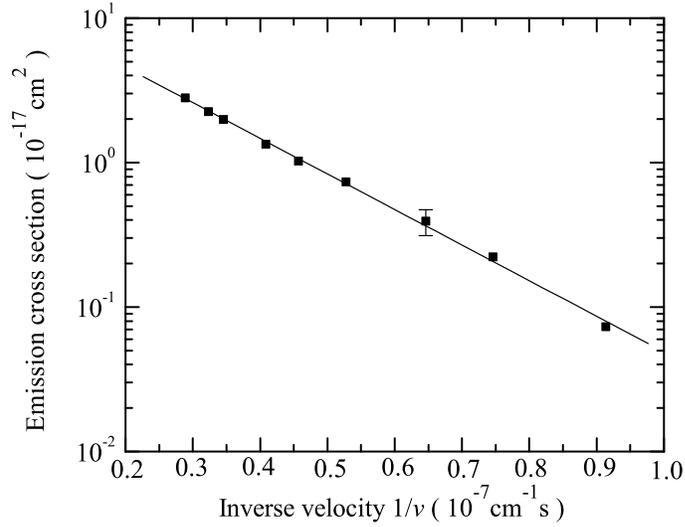} \caption{ Dependence of the emission
cross section for the band (0,0) $\lambda=391.4$ nm of the first-negative band system of N$_{2}^{+}$ ions in O$^{+}$ - N$_{2}$ collision on the inverted
velocity of O$^{+}$ ions.}%
\label{Fig11}%
\end{figure}

\begin{figure}[b]
%\includegraphics[width=7.5cm]{Fig11.eps}
%\caption{(Color online) Dependence of the excitation cross section for the
%band (0,0) $\lambda=391.4$ nm of the first-negative band system of N$_{2}^{+}$ ions
%in O$^{+}$ - N$_{2}$ collision on the inverted velocity of O$^{+}$ ions.}%
\centering
\includegraphics[width=10.0cm]{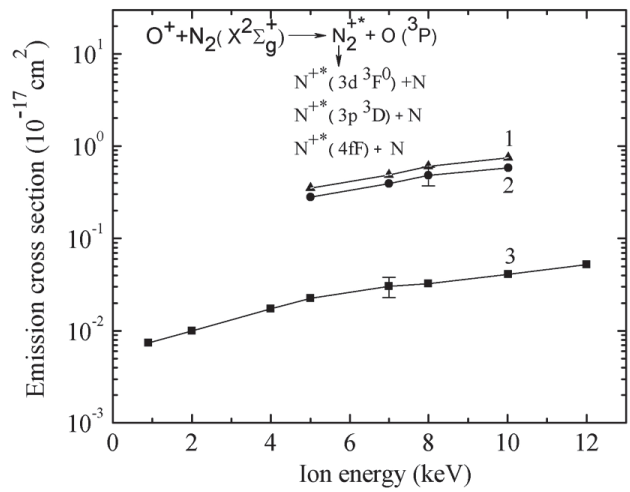} \caption{ Dependence of the emission
cross section for the dissociative product of the N$^{+}$ ion in O$^{+}%
$+N$_{2}$ collisions on the incident energy of O$^{+}$ ions. Curves: 1 -
emission for the line NII $500.5$ nm; 2 - emission for the line NII $567.9$
nm; 3 - emission for the line NII $424.2$ nm.}%
\label{Fig13}%
\end{figure}

\begin{figure}[b]
\centering
\centering
\par
\includegraphics[width=7.9cm]{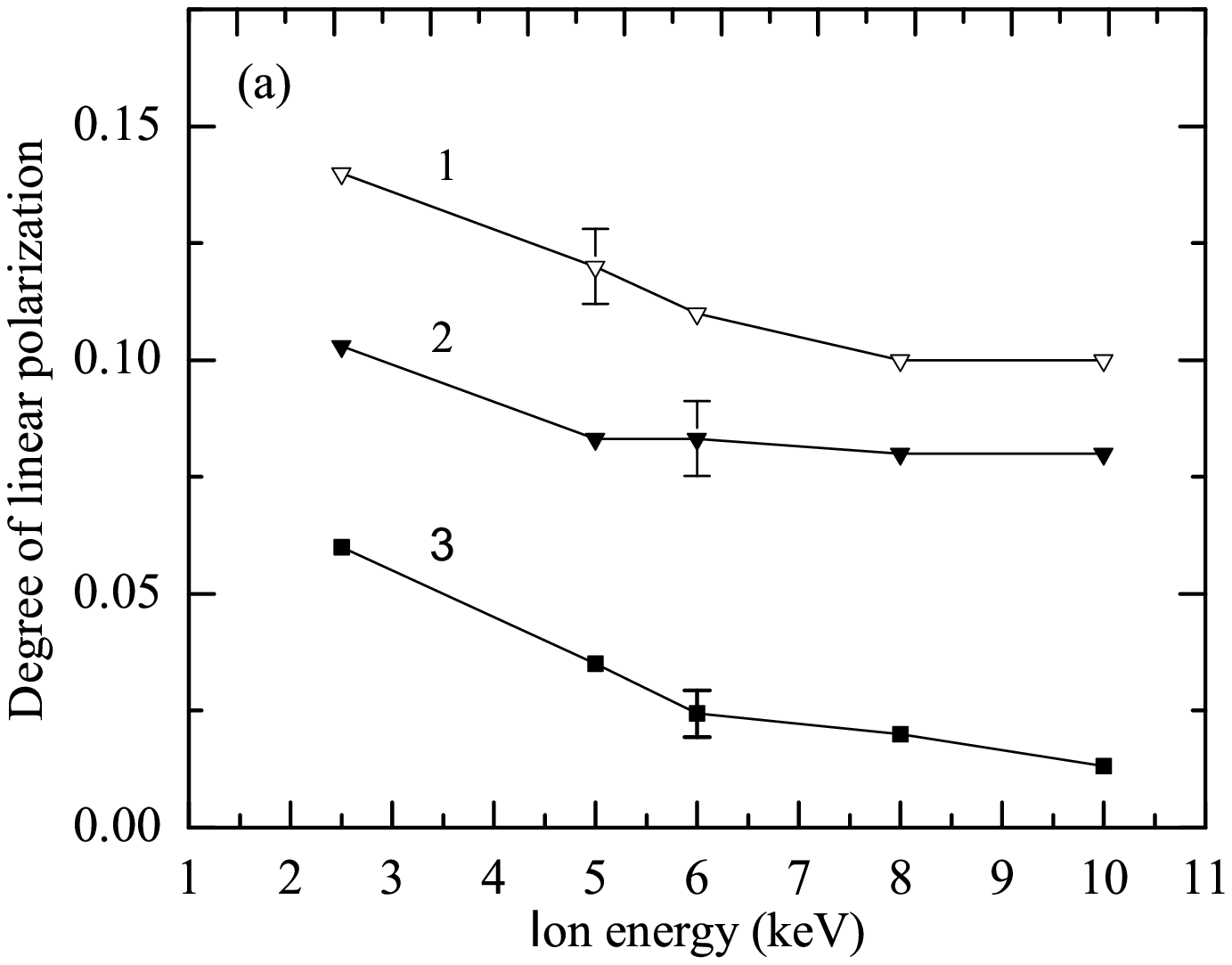}
\includegraphics[width=7.9cm]{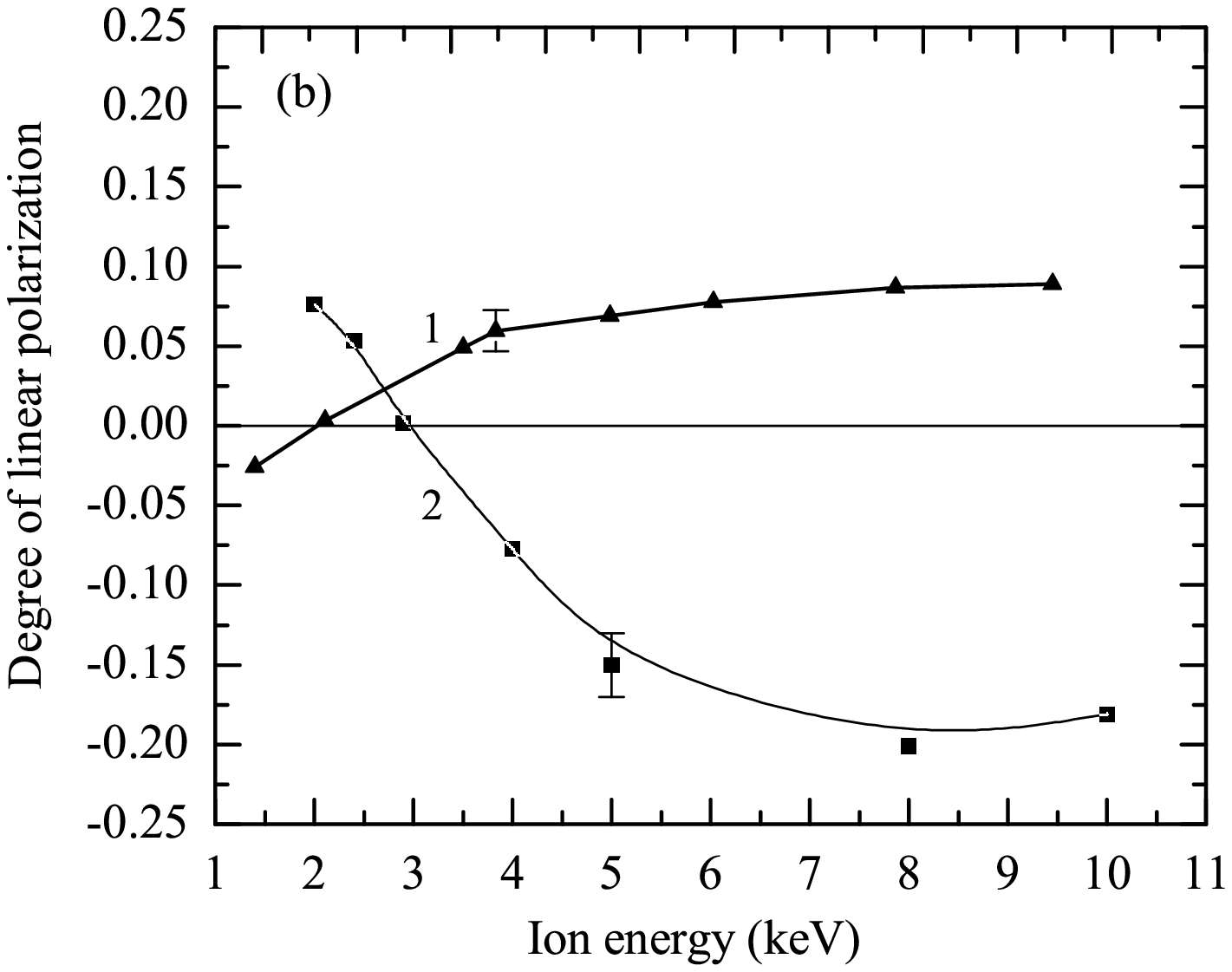}\caption{ Energy dependence of the
degree of linear polarization for the O$^{+}-$N$_{2}$ and He$^{+}-$N$_{2}$
collision systems. (a) The data for the emission of the first negative band
system of the N$_{2}^{+}$ molecular ion in the O$^{+}-$N$_{2}$ collision
system are obtained with different admixture (ground + metastable ) of O$^{+}$
ions. Curves: 1 - the incident O$^{+}$ ions in ground $^{4}S$ state; 2 - the
incident beam contains about 1.4\% of metastable O$^{+}$($^{2}P$) states; 3 -
the incident beam contains about about 7.4\% of metastable O$^{+}$($^{2}P$)
states; (b) The data obtained for He$^{+}-$N$_{2}$ collision. Curves: 1 - the
emission of the first negative band system of the N$_{2}^{+}$ molecular ion; 2
- emission of the excited atomic line $\lambda=389.9$ nm of He($^{3}P$).}%
\label{Fig12}%
\end{figure}

In order to check the influence of the ion source operating parameters and
evaluate the role of metastable ions in emission processes we have made
measurements for the first negative band system of the N$_{2}^{+}$ ion under
different pressure conditions. The dependence of the effective cross section
for the band (0,0) $\lambda=391.4$ nm of N$_{2}^{+}$ on the pressure in the
ion source are presented in figure \ref{Fig10}. The analysis of figure
\ref{Fig10} shows that the intensity of the emission from the decays of the
excited band (0,0) ($\lambda=391.4$ nm) sharply decreases with increasing
pressure in the RF ion source and reaches saturation at $P>$10$^{-1}$ Torr. In
this region the percentage ratio of the metastable and ground states ions does
not change. The abrupt decrease of intensity indicates a decrease of the
relative weight of metastable ions in the ion source. Therefore, in collisions
mainly ions of O$^{+}$ in the ground $^{4}S$ state participate.

To confirm this observation let us consider the dependence of the emission
cross section of the band (0,0) $\lambda=391.4$ nm on the inverted velocity of
the incident ion. This dependence is presented in figure \ref{Fig11}. The
observed linear dependence of the cross section on a semilogarithmic scale
(linear decrease with increase of $1/v$) is nicely described by the relation
$\sigma=Ae^{-a/v},$ where $A$\ and $a$\ are constants. As it is well known
\cite{Nikitin} such a dependence is a peculiar characteristic of inelastic
processes realized in an adiabatic area. From this dependence one can conclude
that the excitation process takes place in one inelastic channel and a
definite portion in it is related solely to the ground state O$^{+}$($^{4}S$)
ions. Thus, we can conclude that it is possible to vary the O$^{+}$ ion beam
content with different internal electronic states (ground and metastable) by
changing the working condition in the ion source.

In our study of O$^{+}-$N$_{2}$ collisions, in addition to molecular bands, we
also observed lines related to the N$^{+^{\ast}}$ ion emissions, which are
formed in the process of dissociative charge exchange processes (4) and (8).
The most clearly identified lines for the dissociative product N$^{+^{\ast}}$,
which have sufficient intensity in the considered energy interval, are
$\lambda=424.2$ nm, $\lambda=500.5$ nm, and $\lambda=567.9$ nm and they are
presented in figure \ref{Fig6}. The dissociation of the N$_{2}(X^{2}\Sigma
_{g}^{+})$ molecule in O$^{+}-$N$_{2}$ collisions leads to the production of
N$^{+^{\ast}}$ions in different excited states. The results of measurements of
the dependence of the emission cross section on the energy of O$^{+}$ ions for
the wavelength $\lambda=500.5$ nm, $\lambda=567.9$ nm and $\lambda=424.2$ nm
are presented in figure \ref{Fig13}. The emission cross sections for the
dissociative product N$^{+^{\ast}}$ increase with increasing the incident ion
energy and show almost a linear dependence on the energy. The cross sections
are measured at a pressure of $2.4\times10^{-2}$ Torr in the ion source. With
increasing pressure inside the source, the ion current at the outlet from the
source drops sharply and, therefore, decreases also in the collision chamber.
As a result the accuracy of the measurements decreases and, therefore, the
sensitivity of the measurement of the emission cross section is significantly
reduced. In particular, with a change in pressure inside the source by an
order of magnitude from $P=2.4\times10^{-2}$ Torr to $P=2.0\times10^{-1}$
Torr, the change in the emission cross section of dissociation products in the
low energy range of $1-2.5$ keV does not exceed 30\% and in the high energy
range of $5-10$ keV $10-15$\%, and the latter one is within the accuracy of
experimental measurements. The analyses of the cross sections for the 3$d$
$^{3}F^{0}\rightarrow$3$p$ $^{3}D$ (the line, NII $500.5$ nm) and 3$p$
$^{3}D\rightarrow3s$ $^{3}P$ (the line NII $567.9$ nm) transitions shows that
while the cross sections are of the same order of magnitude, the increment of
the cross section for the 3$d$ $^{3}F^{0}\rightarrow$3$p$ $^{3}D$ transition
is slightly larger than the one for the 3$p^{3}D\rightarrow3s^{3}P$
transition. Moreover, these cross sections are significantly ($\sim$15 times)
greater than the emission cross section for the $4fF\rightarrow$3$d$ $^{3}D$
(the line NII $424.2$ nm) transition. The similarity of the cross sections for
NII $500.5$ nm and NII $567.9$ nm is probably due to the cascade population of
the $3p$ $^{3}D$ state by the 3$d$ $^{3}F^{0}\rightarrow3p$ $^{3}D$ transition.

\bigskip

To study the effect of different electronic states of ions O$^{+}$ on the
magnitude of cross section for O$^{+}-$N$_{2}$ collision, one can conduct a
study of molecular orientation by measuring the degree of linear polarization
for the emission of excited molecular ion and explore the mechanism of
inelastic processes realized in this collisions. In addition, the polarization
measurements can be also used to separate a pure ground state O$^{+}$($^{4}S$)
ion from the mixture of $^{4}S$ and $^{2}P$ states of the incident beam of
O$^{+}$ ion with high accuracy. The details of this procedure is given in the
Appendix. The energy dependence of the degree of linear polarization for
O$^{+}-$N$_{2}$ and He$^{+}-$N$_{2}$ collision systems is presented in figure
\ref{Fig12}. The data of figure \ref{Fig12}a show the result for O$^{+}%
-$N$_{2}$ collision system and represent the polarization of the emission for
the excited N$_{2}^{+}$ ions obtained for the beam of O$^{+}$ ions in the
ground $^{4}S$ state (curve 1) and 1.4\% and 7.4\% admixture of the metastable
O$^{+}$($^{2}P$) state (curves 2 and 3, respectively). In figure \ref{Fig12}b
a comparison between the polarization for the emission for the same excited
N$_{2}^{+}$ ions (curve 1) obtained for He$^{+}-$N$_{2}$ collision is shown,
while curve 2 represents the degree of linear polarization for the emission of
the excited $^{3}P$ state of He atom. The comparison of the results in figure
\ref{Fig12} (a) and (b) shows that with increasing energy of the incident ions
the degree of linear polarization for the first negative band system of
N$_{2}^{+}$ molecular ion decreases, reaches saturation at high energy and is
always positive. In addition, it is seen that the larger the metastable
$^{2}P$ fraction in the ion beam the smaller the value of polarization. In
spite of that, the portion of the metastable $^{2}P$ state in the total beam
does not exceed 7.4 \% (Fig.\ref{Fig12}a, curve 3), their contribution into
the cross section is determinative. This fact is not surprising because it is
known that the process of charge exchange by oxygen ions in the $^{2}P$
metastable states proceeds with a significantly higher (about two order)
efficiency than by ions when it is in the $^{4}S$ ground state. This also
means, that the process of charge exchange take place at a relatively large
internuclear distance and an influence of the ion electric field on the
electron cloud of the molecule is less pronounced. A relatively small degree
of polarization corresponds to a small electron cloud orientation in space.

To explain the mechanism of population for excited N$_{2}^{+}$ molecular ions
let us consider the results presented in figure \ref{Fig12}b. First of all, we
have to mention that when an emission is polarized then it is anisotropic.
This means that the electronic cloud of excited atomic or molecular particles
should be oriented mainly either perpendicular or parallel to the direction of
the incident ion beam. As seen from figure \ref{Fig12}(b) one can observe a
strong correlation between the degree of polarization for the emission of
excited molecular ions (curve 1) and the emission of excited atomic line
$\lambda=389.9$ nm of He($^{3}P$) (curve 2). In both cases the sign of the
linear polarization changes at $2-3$ keV and at the energy range of $2-10$ keV
the polarization degree of emission for the excited N$_{2}^{+}$ ion reaches
saturation and the sign is positive. The polarization of emission for the
excited He($^{3}P$) atom it also has saturated character but its sign is
negative. The fact, that the sign of the degree of polarization for the band
of molecular N$_{2}^{+}$ ion is positive (opposite to the excited He atomic
line) indicates, that the electronic cloud of excited molecular ion of
N$_{2}^{+}$ is preferably oriented parallel to the direction of the incident
beam and hence a molecular axis is oriented also parallel with respect to the
direction of the incident ion beam. Of course, for a complete understanding of
the excitation mechanism, it would be better to measure the degree of
polarization (or circular polarization) for excited oxygen atomic line,
realized in charge exchange processes O$^{+}$ $+$N$_{2}$ $\rightarrow$
O$^{\ast}+$ \ldots. and apply some theoretical approach for more clarity.
However, in our experiment the value of the cross section for excited oxygen
atomic line was less compared to the band of molecular ion of N$_{2}^{+}$, and
the accuracy of the measurements does not permit to obtain reliable results.
As an alternative way, we checked this opportunity for the excited He atomic
line obtained in collision of He$^{+}$ $+$N$_{2}$ $\rightarrow$ He$^{\ast}+$
\ldots\ hoping that an obtained opposite result will serve for establishing
the excitation mechanism of N$_{2}^{+}$ molecular ion.

\section{CONCLUSIONS}

Optical spectroscopy, in combination with an intense RF ion source, was
applied
%to explore the pressure condition in the ion source and
to perform emission measurements for molecular N$_{2}^{+}$ and atomic N$^{+}$
ions in collisions of O$^{+}$ ($^{4}S$, $^{2}D,^{2}P$) with N$_{2}$ molecules. The measurements are
performed with the sufficiently high spectral resolution, which
allows to distinguish the emission channels.
The emission cross section of N$_{2}^{+}$ ion for the (0,0), (0,1) and (1,2)
bands systems in collision of mixed states of O$^{+}$ ions beam with nitrogen
molecules have been measured. The ratio of the intensities for these bands was
established \ as $10:3:1$. The exact edge of the pressure condition in the RF
source was revealed which enabled us to distinguish ground state and
metastable states ions in the primary beam. The presence of ions in the
metastable states of $^{2}P$ and $^{2}D$ was monitored by measurement of the
energy dependence of \ the emission cross section of the first negative band
system (0, 0) $\lambda=391.4$ nm of N$_{2}^{+}$ ion for different pressure
condition in the RF source. An influence of the pressure condition in the RF
source on the formation of certain O$^{+}(^{2}P)$ states was established. Its
percentage value for a pressure in the RF source of $P=2.4\times10^{-2}$,
$8\times10^{-2}$ and $2\times10^{-1}$ Torr was determined to be 7.4\%, 3.4\%
and 1.4\%, respectively. A small fraction of metastable $^{2}D$ ions in the
ion source compare to $^{2}P$ and $^{4}S$ states was identified by
measurements of the Meinel band system (3,0) $\lambda=687.4$ nm and (4,1)
$\lambda=703.7$ nm.

The absolute emission cross section for (0,0) $\lambda=391.4$ nm band of the first-negative band system of the N$_{2}^{+}$ by O$^{+}$ ions in the ground $^{4}S$
and metastable $^{2}P$ state was obtained in the energy interval $1-10$ keV.
The influence of ions, having different electronic states (ground and
metastable) on the magnitude of the cross section was established. It was
found that metastable $^{2}P$ state ions indeed enhance the excitation cross
section compared to those measured for the ground O$^{+}(^{4}S)$ state ions.
While the emission cross section obtained for the O$^{+}(^{4}S)$ state ions
sharply increases in all energy interval, the emission cross section for the
incident O$^{+}$ ions in $^{2}P$ state is larger by two orders of magnitude at
1 keV and 10 time larger at 10 keV than that one for the O$^{+}(^{4}S)$
incident ions.

In the case of collisions of O$^{+}$ with nitrogen molecules in the incident
beam in emission processes, we used a filtration chamber, as in the earlier
works, to assess the role of the metastable state O$^{+}(^{2}D$, $^{2}P$).
However, the originality of our work lies in the fact that ion filtration
occurs in the RF ion source. The latter allows us to change the pressure and
the mixture of gases inside the source, the power deposited inside the
discharge, and other conditions which have an influence on formation of the
incident ions beam.

The cross sections for the N$^{+^{\ast}}$ions emissions in the dissociative
charge exchange processes increase with increasing incident ion energy and
show almost a linear dependence on the energy. The cross sections for the 3$d$
$^{3}F^{0}\rightarrow$3$p$ $^{3}D$ and 3$p$ $^{3}D\rightarrow3s$ $^{3}P$
transitions are of same order of magnitude but significantly greater than the
emission cross section for the $4fF\rightarrow$3$d$ $^{3}D$ transition. The
similarity of the cross sections for the 3$d$ $^{3}F^{0}\rightarrow$3$p$
$^{3}D$ and 3$p$ $^{3}D\rightarrow3s$ $^{3}P$ transitions is probably due to
the cascade population of the $3p$ $^{3}D$ state by the 3$d$ $^{3}%
F^{0}\rightarrow3p$ $^{3}D$ transition.

The mechanism of the processes realized during collisions of ground and
metastable oxygen ions on molecular nitrogen have been established by the
study of molecular orientation effect and measurement the degree of linear
polarization for emission of excited molecular ions. It was demonstrated that
for the O$^{+}-$N$_{2}$ collision system the degree of linear polarization by
the metastable O$^{+}(^{2}P)$ ions is less compared to those that are in the
ground O$^{+}(^{4}S)$ state and the sign of the degree of linear polarization
of excited molecular ions does not change.

This is the first measurements of the energy dependence of the emission cross
section of the band (0,0) $\lambda=391.4$ nm of the first-negative band system of
the N$_{2}^{+}$ in a wide energy range of $1-10$ keV of the incident
O$^{+}(^{4}S)$ and O$^{+}(^{2}P)$ ions, as well as the degree of linear
polarization of emission was measured for the first time.

\bigskip

\appendix

\section{\label{Polarization}Polarization}

The polarization of radiation can be written as:%

\begin{equation}
\Pi=\frac{I_{||}-I_{\perp}}{I_{||}+I_{\perp}}, \label{Poldef}%
\end{equation}
where $I_{||}$ and $I_{\perp}$ intensities of emission along\ and
perpendicular to the ion beam, respectively.

\bigskip Polarization of the emission emerging from excited $^{3}P$ state of
helium is connected to the relative populations of m$_{l}=0$ and m$_{l}=\pm1$
sublevels. The m$_{l}=0$ correspond to the redistribution of electron cloud to
the direction of ion beam and m$_{l}=\pm1$ to the perpendicular, respectively.
Expression for the first Stock's parameter has been derived on the basis of
general approach developed by Macek and Jaecks in \cite{Macek}. Here is
presented only the final formula for the linear polarization:%

\begin{equation}
\Pi=\frac{I_{||}-I_{\perp}}{I_{||}+I_{\perp}}=\frac{15\left[  \sigma
(m_{l}=0)-\sigma(m_{l}=\pm1)\right]  }{41\left[  \sigma(m_{l}=0)+67\sigma
(m_{l}=\pm1)\right]  }, \label{Polar}%
\end{equation}
Taking into account that experimentally observed value of $\Pi$ is 20\% (sign
is negative), one obtains from (\ref{Polar}) that $\sigma(m_{l}=\pm
1)/\sigma(m_{l}=0)\approx15$. Such a great value of this ratio indicates that
$m_{l}=\pm1$ sublevels of the excited helium atom line are preferably
populated. This means that the electron density formed in He$^{\ast}$ during
the collision is oriented perpendicularly with respect to the incident beam direction.

In this respect, it should be bear in mind that the ion current is
two-component ions in the ground $^{4}S$ and metastable $^{2}P$ states,
respectively. We have already determined their relative contribution in the
total current, which depends on the pressure in the ion source. To perform the
calculations, we considered the data for two different pressures\ $P_{1}%
=2.4\times10^{-2}$ Torr and\ $P_{3}=20.0\times10^{-2}$ Torr in the RF ion
source. Our calculations above show that for these pressures $f_{0}=0.926$
$\ $and $f_{2}=0.074,$ and $f_{0}=0.986$ and $f_{2}=0.014,$ respectively.

For a given $P$ pressure, equations for $I_{||}$ and $I\perp$ components of
the intensities can be written as%

\begin{equation}
I_{||}(P)=f_{0}(P)I_{||}(^{4}S)+f_{2}(P)I_{||}(^{2}P),\text{ \ \ }I_{\perp
}(P)=f_{0}(P)I_{\perp}(^{4}S)+f_{2}(P)I_{\perp}(^{2}P).
\end{equation}
The magnitude of the polarization is determined for the given energy and for
the given state of the ions. It will be recorded as follows:%

\begin{equation}
\Pi(^{4}S)=\frac{I_{||}(^{4}S)-I_{\perp}(^{4}S)}{I_{||}(^{4}S)+I\perp(^{4}%
S)},\text{ \ \ \ }\Pi(^{2}P)=\frac{I_{||}(^{2}P)-I_{\perp}(^{2}P)}{I_{||}%
(^{2}P)+I_{\perp}(^{2}P)},
\end{equation}
While the experimental value is determine by Eq. (\ref{Poldef})%

\begin{align}
\Pi(P_{1})  &  =\frac{f_{0}(P_{1})\left[  I_{||}(^{4}S)-I_{\perp}%
(^{4}S)\right]  +f_{2}(P_{1})\left[  I_{||}(^{2}P)-I_{\perp}(^{2}P)\right]
}{f_{0}(P_{1})\left[  I_{||}(^{4}S)+I_{\perp}(^{4}S)\right]  +f_{2}%
(P_{1})\left[  I_{||}(^{2}P)+I_{\perp}(^{2}P)\right]  },\\
\text{ }\Pi(P_{3})  &  =\frac{f_{0}(P_{3})\left[  I_{||}(^{4}S)-I_{\perp}%
(^{4}S)\right]  +f_{2}(P_{3})\left[  I_{||}(^{2}P)-I_{\perp}(^{2}P)\right]
}{f_{0}(P_{3})\left[  I_{||}(^{4}S)+I_{\perp}(^{4}S)\right]  +f_{2}%
(P_{3})\left[  I_{||}(^{2}P)+I_{\perp}(^{2}P)\right]  },\text{\ }%
\end{align}
In the given equations, we can convert the effective cross section for the
fixed energy and the different pressure $P_{1}$ and $P_{3}$ in the ion source.
It is clear that the intensity of emission (standardized ion current and
pressure in the collision chamber) measured in the experiment differs from the
effective cross section with a constant coefficient $a$. That is,%

\begin{align}
\sigma_{eff}(P_{1})  &  =a\left[  f_{0}(P_{1})I_{||}(^{4}S)+f_{2}(P_{1}%
)I_{||}(^{2}P)+f_{0}(P_{1})I_{\perp}(^{4}S)+f_{2}(P_{1})I_{\perp}%
(^{2}P)\right]  \text{,}\\
\sigma_{eff}(P_{3})  &  =a\left[  f_{0}(P_{3})I_{||}(^{4}S)+f_{2}(P_{3}%
)I_{||}(^{2}P)+f_{0}(P_{3})I_{\perp}(^{4}S)+f_{2}(P_{3})I_{\perp}%
(^{2}P)\right]
\end{align}
where $a$ is the constant of the absolute calibration in our conditions.
Similarly $\sigma_{eff}=a\left[  I_{||}(^{4}S)+I_{\perp}(^{4}S)\right]  $ and
$\sigma_{eff}=a\left[  I_{||}(^{2}P)+I_{\perp}(^{2}P)\right]  $ represent the
absolute values {}{}of the effective cross section for the given $E$ energy
for the ground $^{4}S$ and metastable $^{2}P$ states, respectively.

Finally we have a system of equations:%

\begin{align}
\Pi(P_{1})  &  =\frac{f_{0}(P_{1})x+f_{2}(P_{1})y}{\sigma_{eff}(P_{1})},\\
\text{ }\Pi(P_{3})  &  =\frac{f_{0}(P_{3})x+f_{2}(P_{3})y}{\sigma_{eff}%
(P_{3})},\text{\ }%
\end{align}
where $x=a\left[  I_{||}(^{4}S)-I_{\perp}(^{4}S)\right]  $ and $y=a\left[
I_{||}(^{2}P)-I_{\perp}(^{2}P)\right]  $.

We solve the system of equations and find the values {}{}of $x$ and $y$. Then
one can determine the polarization as $\Pi(^{4}S)=\frac{x}{\sigma_{0}(E)}$ and
$\Pi(^{2}P)=\frac{y}{\sigma_{2}(E)}$, where $\sigma_{0}(E)$ and $\sigma
_{2}(E)$ are the absolute emission cross sections for the processes induced by
O$^{+}(^{4}S$) and O$^{+}(^{2}P$) ions, respectively, at the energy $E$. We
performed this procedure at all energies and determined the energy dependence
of the degree of linear polarization of the nitrogen molecular ion emission.
In figure \ref{Fig12}a curve 1 shows the results for the pure $^{4}S$ state
only. The polarization measurement accuracy for the $^{2}P$ is much less than
for $^{4}S$.

\end{document}